# A discussion of measuring the top-1% most-highly cited publications: Quality and impact of Chinese papers

Caroline S. Wagner,[a]*Lin Zhang,[b] and Loet Leydesdorff [c]

**Abstract**

The top-1% most-highly-cited articles are watched closely as the vanguards of the sciences. Using Web of Sciencee data, one can find that China had overtaken the USA in the relative participation in the top-1% (PP-top1%) in 2019, after outcompeting the EU on this indicator in 2015. However, this finding contrasts with repeated reports of Western agencies that the quality of China's output in science is lagging other advanced nations, even as it has caught up in numbers of articles. The difference between the results presented here and the previous results depends mainly upon field normalizations, which classify source journals by discipline. Average citation rates of these subsets are commonly used as a baseline so that one can compare among disciplines. However, the expected value of the top-1% of a sample of $N$ papers is $N/100$, *ceteris paribus*. Using the *average* citation rates as expected values, errors are introduced by (1) using the mean of highly skewed distributions and (2) a specious precision in the delineations of the subsets. Classifications can be used for the decomposition, but not for the normalization. When the data is thus decomposed, the USA ranks ahead of China in biomedical fields such as virology. Although the number of papers is smaller, China outperforms the US in the field of Business and Finance (in the *Social Sciences Citation Index*; $p<.05$). Using percentile ranks, subsets other than indexing-based classifications can be tested for the statistical significance of differences among them.

**Keywords:** citation analysis; national standing; international cooperation; science; technology; United States; China

[a] * Corresponding author; John Glenn College of Public Affairs, The Ohio State University, Columbus, Ohio, USA. ORCID: 0000-0002-1724-8489; wagner.911@osu.edu
[b] School of information management, Wuhan university, Wuhan, China. ORCID: 0000-0003-0526-9677; linzhang1117@whu.edu.cn

[c] Amsterdam School of Communication Research (ASCoR), University of Amsterdam, PO Box 15793, 1001 NG Amsterdam, The Netherlands; ORCID: 0000-0002-7835-3098; loet@leydesdorff.net

# 1. Introduction

Indicators of national standing in science and technology influence policy and investment, but choices about measurement present challenges related to comparability of output. Schubert & Braun (1986) noted that pu[1]blication and citation counts used to compare nations should only be applied after proper standardization and normalization. However, the question of 'proper' standardization and normalization is highly technical. Choice of measures around quality and impact differ, and, for each, data normalization and customization can change the outcome, but the embedded choices are unknown to most users. Comparisons require a reference standard that accounts for differences in performance and impact. In this paper we discuss the influence of different choices for selecting the top 1% most highly cited articles. We use recent Chinese articles as an example.

We chose China because of a discrepancy for 2019 between one measure showing Chinese-authored work to be at the forefront of a quality indicator, while another report showed Chinese work as lagging the leading edge. Using the data for China makes an interesting case because that country's scholarly publications have been growing spectacularly since the mid-1990s (e.g., Jin & Rousseau, 2004; King, 2004; Moed, 2002; Zhou & Leydesdorff, 2006). A number of scholars and leading policy organizations have watched this development with great interest. However, services responsible for comparing national science and technology outputs—such as the National Science Board (NSB)[4] of the United States—have reported that the quality of China's output is lagging (NSB, 2020) the United States and European nations. In *The State of U.S. Science & Engineering* (NSB, 2020), for example, the large gap between the US and China in the top-1% most-highly-cited was displayed with the following figure showing the United States outperforming other nations in the top 1% most-highly cited (Figure 22, https://ncses.nsf.gov/pubs/nsb20201/global-science-and-technology-capabilities).

---

[1] The National Science Board of the United States establishes the policies of the National Science Foundation and serves as advisor to Congress and the President. NSB's biennial report—*Science and Engineering Indicators (SEI)*—provides comprehensive information on the nation's S&E.

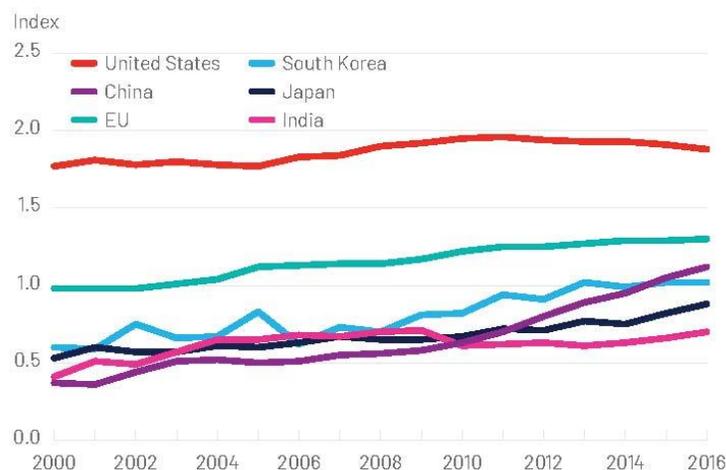

Figure 22. Top 1% cited article index, by selected region, country, or economy: 2000-16

Note: Citation counts for a year are the number of citations in the peer-reviewed literature for articles published in that year. At least 2 years of data after publication are needed for a meaningful measure. See p. 22. SOURCE: NCSES, special tabulations (2019) of Elsevier's Scopus database. *Indicators 2020: Publication Output*

**Figure 1**: Top-1% most highly cited papers as a percentage of national outputs after normalization. Source: NSB, 2020, p. V-12, Fig. 22.

The National Science Foundation (NSF) considered these normalized values as an index that is similar to a standardized score, calculating the measure by comparing "the representation of its articles among the world's top 1% of cited articles, normalized to account for the size of each country's pool of publications. This normalized value is referred to as an index and is similar to a standardized score." A recent European report used similar methods (Jonkers *et al*., 2021), and found that China had overtaken the EU in 2016 when using PP-top1% (that is, the relative participation in the top-1%) for the measurement. They also reported that China still lagged the United States in the top-1%.

In the technical documentation of the data-processing used for preparing the *Science and Engineering Indicators* (NSB, 2020), Coté *et al.* (2019) describe the normalization as follows:

"For a paper in a given subfield (based on the classification of journals described previously in this section) and publication year, the citation count is then divided by the average count of all papers in the relevant subfield (e.g., astronomy & astrophysics) and publication year to obtain an RC." (p. 29; RC means "relative citation score," that is, relative to the mean). However, *RC* is essentially the same as the *Mean Normalized Citation Score* (MNCS) of CWTS in Leiden (Waltman *et al.*, 2011; cf. Tijssen et al., 2002, Table 1, p.388, Note b). These are different names for the standard normalization against averages of the citation distributions as the expected values. Elsevier's *Field-Weighted Citation Impact* (FWCI) is similarly defined as "the ratio of the total citations received, and the total citations that would be expected based on the *average* of the citation rates in subject fields" (e.g., Scopus, 2000).

We had reason to question this approach when we noted that China was leading in citation rates in several fields, but this did not seem to be reflected in the aggregate measure shown in the NSF report. Drawing data from the "Flagship Collection"[2] of the Web of Science, we attempted to replicate Figure 1, but we changed the approach and we used a different measurement tool, which we explain below. We found that China surpassed the United States in 2019 in top 1% most highly cited (Figure 2). The difference between PP-top1% of China and the United States in 2019 is not statistically significant ($z = 0.80$), whereas the differences between the United States and China were statistically significant when the United States outperformed China in the previous years.[3] The different results are not simply a matter of different assumptions having different outcomes but about choosing accurate measurement tools. Ours is not just another representation; but could be seen as a methodological improvement.

---

[2] The "Flagship Collection" of WoS includes the Science Citation Index-Expanded (SCIE), the Social Sciences Citation Index, (SSCI), and the Arts & Humanities Citation Index (A&HCI), and the Emerging Sources Citation Index.

[3] Interestingly, since 2017, the differences between the EU with or without the UK are not statistically significant ($z \leq 1.96$; $p > .05$).

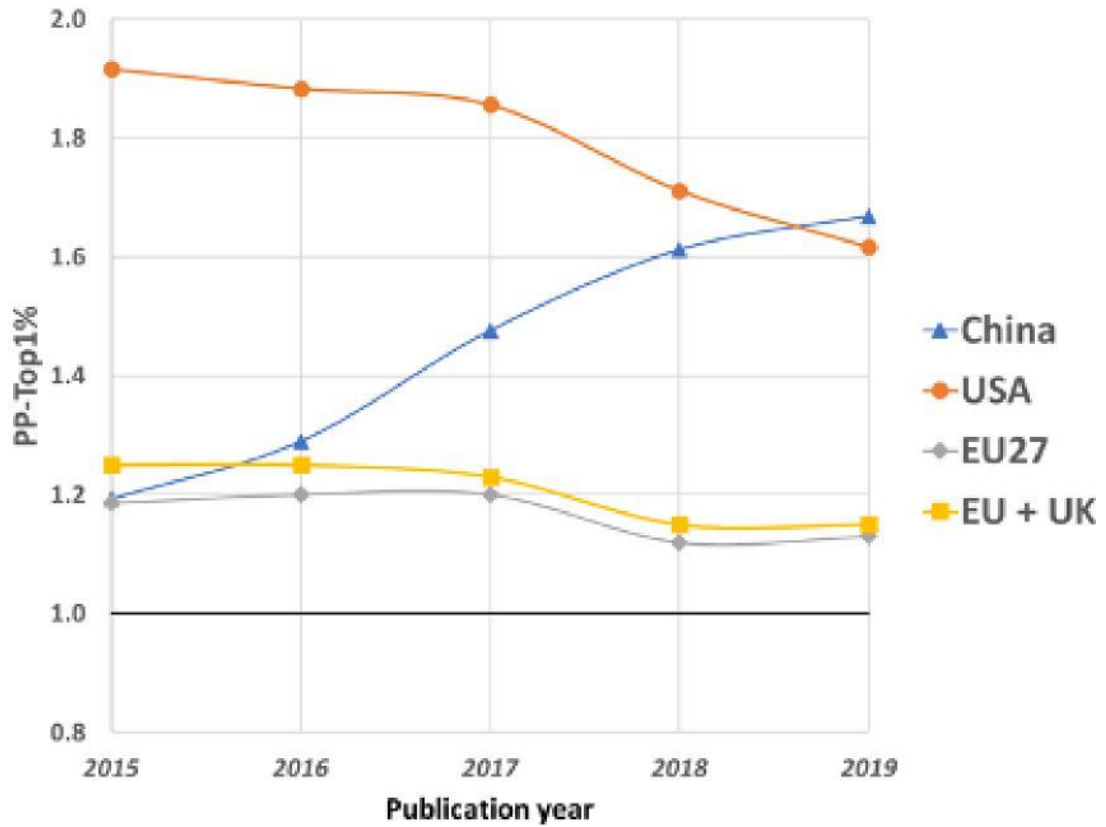

**Figure 2**. Percentage participation in the top-1% (PP-Top1%) most-highly-cited publications (worldwide) by papers with an address in the United States, China, and the EU (with and without the UK); articles, reviews, and letters; retrieval from the Web of Sciencee "Flagship" Collection on March 6, 2021.

## 2. Methods

We used Ahlgren *et al.*'s (2014) methodology for determining the percentage of publications of a country in the top-1% segment (PP-top1%) using the online version of WoS on March 6, 2021. The method is used in the following way:

1. First, we retrieved the top-1% of the set of letters (as they are important in some leading journals such as Nature (Braun et al., 1989; cf. Narin, 1976)), articles, and reviews. Hereafter, we use the term "articles." We collected articles published in 2019 by searching with "(py = 2019) *AND* DOCUMENT TYPES**:** (Article OR Letter OR Review) Indexes=SCI-EXPANDED, SSCI, A&HCI." The resulting number for the global system in 2019 was 2,041,287 at the retrieval date (March 6, 2021; see Table 1).

2. After sorting the retrieval in descending order by citations, the threshold citation rate of the top-1% class (worldwide) is 20,413, or, one percent of 2,041,287. Record number 20,413 had 38 citations on March 6, 2021. Thus, all papers published in 2019 with 38 or more citations belong to the top-1% class of the database in 2019.
3. The above searches can be limited to a domain—for example, the United States—using "(CU = USA AND py = 2019)". Among the 492,448 articles with at least one American address, 7,959 had 38 or more citations (see columns b and c of Table 1). This is the observed number of US publications in the top-1% segment. Without prior knowledge of the American system, the expected number of publications in the top-1% layer is—for analytical reasons—1% of 492,448 = 4,924.

**Table 1.** Descriptive statistics of publication rates in the top-1% most-highly-cited segment for China, the United States, and the EU (with and without the UK); retrieved on March 6, 2021. ($N$ is the number of articles, reviews, and letters; *P-Top 1%* is the *number* of publications participating in the top-1% of the respective set; *PP-Top-1%* is the *percentage* of publications participating in the top-1% of the set.)

|      | $N$       | *P-Top 1%* | *PP-Top 1%* |
|------|-----------|------------|-------------|
| (a)  | (b)       | (c)        | (d)         |
| China | 504,695  | 8,422      | 1.67        |
| USA   | 492,448  | 7,959      | 1.62        |
| EU-27 | 536,932  | 6,074      | 1.13        |
| EU + UK | 639,217 | 7,337    | 1.15        |
| World | 2,041,287 | 20,413    | 1.00        |

The PP-top1%—the percentage of participation in the top-1%—in our approach is defined as the ratio between observed and expected numbers of top-1% papers. For example: the PP-top1% value for the United States in 2019 is the observed value of 7,959, divided by the expected value of 4,924 = 1.62 (in column d). Using the *z*-test (Leydesdorff & Bornmann, 2012), one can specify whether differences between observed and expected values are statistically significant. PP-top1% is based on percentages and thus presents a size-independent measure of quality.

## 3. Explaining the different approaches

The results presented in Figure 2 are different from those provided in the *Science and Engineering Indicators* 2020 and others for two major reasons. First, we did not normalize for fields, although we are aware that this is common practice in bibliometrics. Our method uses percentile ranks (Leydesdorff & Bornmann, 2011; Bornmann & Mutz, 2011). *Ceteris paribus*, one can expect that 1% of a random sample will contain 1% of papers belonging to the top-1% (or analogously the bottom-1%). The deviation from this expectation can be statistically tested for its significance. Second, we do not use averages across fields, since the data are highly skewed. In summary, we see two problems with the existing methods: 1) binning articles based upon the journal's field introduces a fallacy and treats journal contents as homogenous; and 2) citation normalizations use the means of highly skewed distributions, which is generally considered invalid in statistics.

Let us take the first problem: that of the field classification of articles based upon the journal. The classification and binning of articles, reviews and letters into fields is an artifact of the earliest days of the Web of Science, when Subject Categories (WCs) were created (during the 1960s; see Bensman, 2007). Pudovkin & Garfield (2002, p. 1013, n. 11) described the procedures used by the Institute for Scientific Information (ISI)[4] in establishing journal categories as follows:

> *These procedures are followed by the ISI editorial group in charge of journal selection and are similar to those used for the SCI and Current Contents® journal categories. This method is "heuristic" in that the categories have been developed by manual methods started over 40 years ago. Once the categories were established, new journals were assigned one at a time. Each decision was based upon a visual examination of all relevant citation data. As categories grew, subdivisions were established. Among other tools used to make individual journal assignments, the Hayne-Coulson algorithm is used. The algorithm has never been published. It treats any designated group of journals as one macrojournal and produces a combined printout of cited and citing journal data.*

---

[4] The Institute for Scientific Information (ISI) in Philadelphia (PA) was the owner and producer of the *Science Citation Indexes* at the time.

ISI/Clarivate has used, developed, and expanded the classification over time. However, the historical backbone of the classification has remained constant even as new fields were added and older fields change (Boyack *et al*., 2005; Leydesdorff & Bornmann, 2016; cf. Klavans & Boyack, 2009; Archambeau *et al*., 2011). Moreover, different ontologies have been developed over time (Glänzel & Schubert, 2003), making it difficult to compare classifications across studies and databases. Science-Metrix, for example, uses an inhouse taxonomy of 14 disciplines which are divided into subfields, which can be reclassified (Coté et al., 2019). Propriety classifications make replication difficult.

Field attributions at the journal-level miss the point that journals can contain articles that represent contributions to and from different fields. Rafols & Leydesdorff (2009) distinguished between indexer-based classifications and algorithm-based decompositions, but, since there is no consensus on a disciplinary ontology, the choice of categories is driven by institutional interests (Griliches, 1994) or the needs of the analyst. For example, the Expertise Center for Research Management and Development Monitoring (ECOOM) in Leuven, Belgium, developed an indexer-based scheme with fewer categories than ISI's 254 WCs (Glänzel & Schubert, 2003); the Centre for Science and Technology Studies (CWTS) in Leiden uses more than 4000 micro-specialties on the basis of an algorithmic decomposition (Waltman & van Eck, 2012). The micro-specialties are aggregated into five broad fields. The latter system is integrated into VOSViewer for the purpose of visualizations (Waltman, van Eck, & Noyons, 2010). Each of these categorizations can be useful in their own way, but a multitude of possible ontologies complicates comparisons.

Furthermore, field-attributions are particularly unreliable in cases of multi- or interdisciplinarity. Papers published in general-science journals such as *Nature* or *Science*, or those in inter- or multidisciplinary journals present an additional challenge around field-level classification. Papers in general-science journals are sometimes reclassified at the article level, but without reference to a single ontology (Milojević, 2000).

The process of attributing higher-level categories (at the journal level) to lower-level units of analysis (at the article level) is known as an "ecological fallacy" (Robinson, 1980) in that inferences about the nature of individuals cannot be deduced from inferences about the group to which those individuals belong. This ecological fallacy appears to be at work in binning articles into fields based upon the journal's titular discipline.

Let us turn to the second problem, that of averaging citations at the field level. It is well known that citations are not normally distributed. When data are not normally distributed, averages are invalid. Generally, in these cases, non-parametric statistics—in which the data are not assumed to come from prescribed models—are more reliable. Field normalization, for example, brings the average value as a constant into the denominator for each field. These 'constants' function as weights, and they confound the analysis. When the *average* values for fields of science are used as the expected values, systematic error terms can be introduced by (1) using the means of highly skewed distributions and (2) applying the delineations of subsets by field. In our opinion, these problems can be avoided by applying non-parametric statistics. In this paper, we use the chi-square statistics and not the t-test between means. However, Figure 1 above was based on both sources of problems: the data were field normalized using an *ad hoc* classification, even though field classification is suspect, and averages were used to normalize citations, even though averaging along a non-normal distribution is invalid.

In response to a conference presentation of this paper, David Campbell (Chief Scientist of Science-Metrix; personal communication, 15 July 2021) emphasized that, without field normalization, fields with high citing norms will be favored. In our opinion, pulling highly cited articles without reference to fields (while still attributing articles to the source country) may give a better indication of national standing. We examined ISI/Clarivate's method of determining relative citation and we found that they divide the percentile rank of each paper by the *average* percentile of 22 "broad categories" distinguished in InCites ($c_i / \sum_{i=1..n} c_i$ or RC) to normalize, as Science-Metrix does, InCites uses the average of the percentile distribution [($p_r / \sum_{i,r=0...99}^{99} p_i$)].

In this case, the top-1% is thus normalized in terms of the 22 broad categories of ISI's

Essential Science Indicators (ESI); papers in this subset are assigned to the so-called "ESI refined top-cited papers."

We use the same values in the numerator, but do not divide by the average percentile or the average of the citation distribution in the respective classes. The average percentile or citation is a constant for each (sub)field *f*. This constant can also be considered as a weight ($1/f$) in the impact calculations. As in the case of relative citation rate (RCR), mean-normalized citation scores (MNCS), and FWCI, the normalization does not affect the rank-order within a category, but papers in different categories are weighted differently.

Instead of dividing by a mean value of a field, we first aggregate papers by citations by nation. The value of PP-top1% is then based upon counting: each paper in the highly cited set contributes independently (not weighted by field) and is normalized against the expected citation rate in its percentile class rather than its field. The groupings can be changed because the numerators are "1" for all papers. For example, one can use all papers in journals with a specific WC attributed to the source journals; but one can also use different sets. Thus, the sets under study can be decomposed, reaggregated, and changed depending on the research question.

Our approach is done to avoid introducing one weight or another on the basis of a prior classification scheme, but to weight all papers equally as ones. We do not *ex ante* correct the lower impact of, for example, the social sciences, in terms of citation rates by multiplying with a group-specific constant derived at the aggregated level, as this can lead to the aforementioned fallacy; but one can distinguish between quality and impact. Even if a contribution from the social sciences is of high quality, it may have low impact. PP-top1% is a quality indicator and not an impact indicator. Impact cannot be size-independent, and one cannot disregard less-cited papers when computing impact. That this feels as "unfair" introduces a normative concern; however, the database can remain neutral.

We accessed a table provided by Clarivate/ISI in ESI at the WoS interface with the values thus normalized for the 254 Web of Sciencee WCs in the core collection.[8] These "ESI refined top-1% cited papers" allow us to replicate the analysis above. In Figure 3, the results are compared with the values provided in Figure 2.

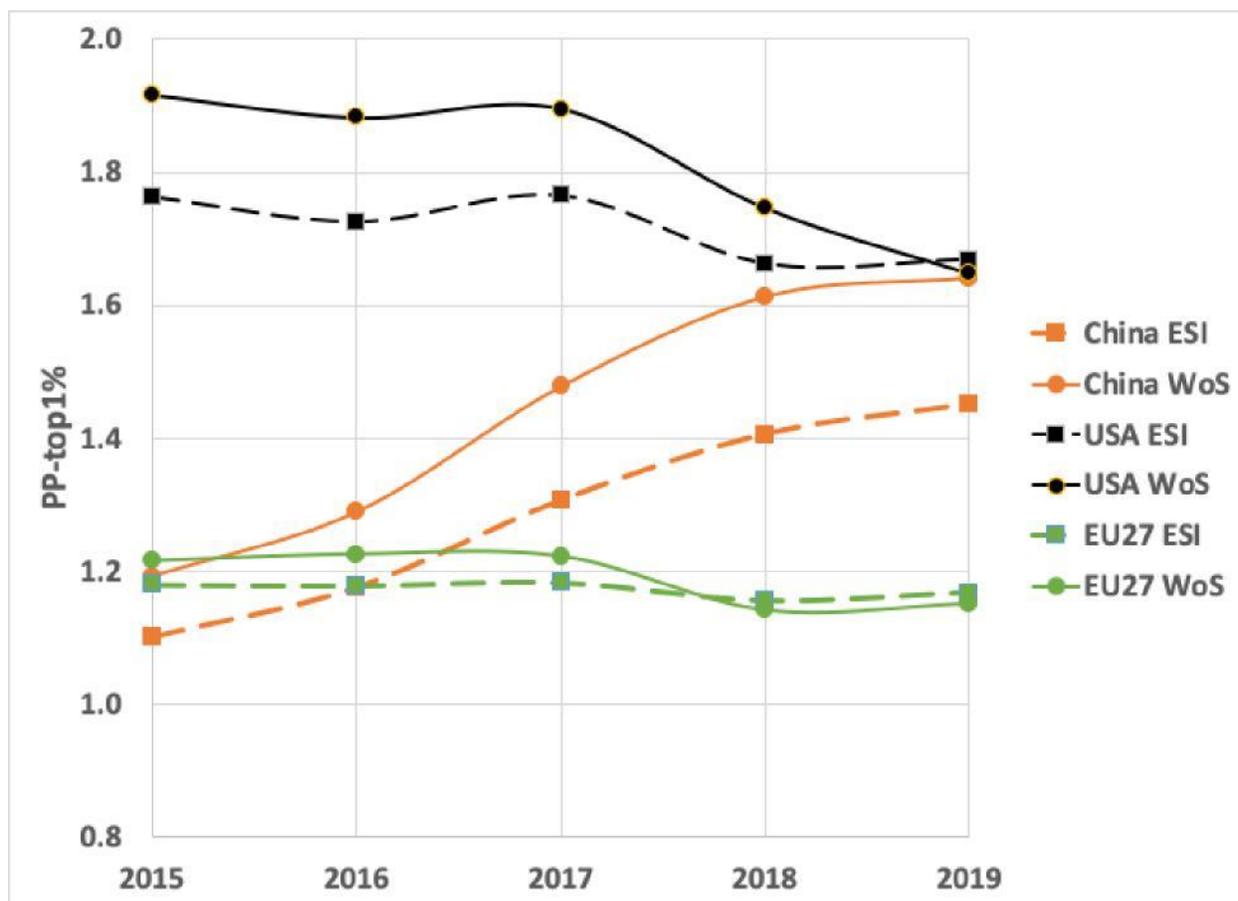

**Figure 3**. Differences between ESI-normalized PP-Top1% values (dashed lines) and data from Figure 2 for papers with an address in the United States, China, and the EU-27; articles, reviews, and letters; retrieval from the Web of Science Core Collection on March 6, 2021.

The "ESI-refined" papers have lower scores than the WoS-based ones since, in a right-skewed distribution, the mean is larger than the median and thus *a fortiori* larger than the first percentile. With a larger value in the denominator, the "refined" calculation depresses the resulting values. Since the mean of the citations is larger than the median of the percentile distribution in a right-skewed distribution, the underrepresentation is larger in the case of Figure 1 (using Science-Metrix's methodology) than in Figure 3 (using Clarivate's methodology).

Normative concerns and theoretical considerations may lead an analyst to choose collections of journals, keywords, or classifications. In Figure 4, for example, we added the PP-top1% values compared to international collaborations based on considering China, the United States, and EU27 as the main contributors at the global level. This research question is different from, but akin to, the research questions in this paper; but the methodology is the same.

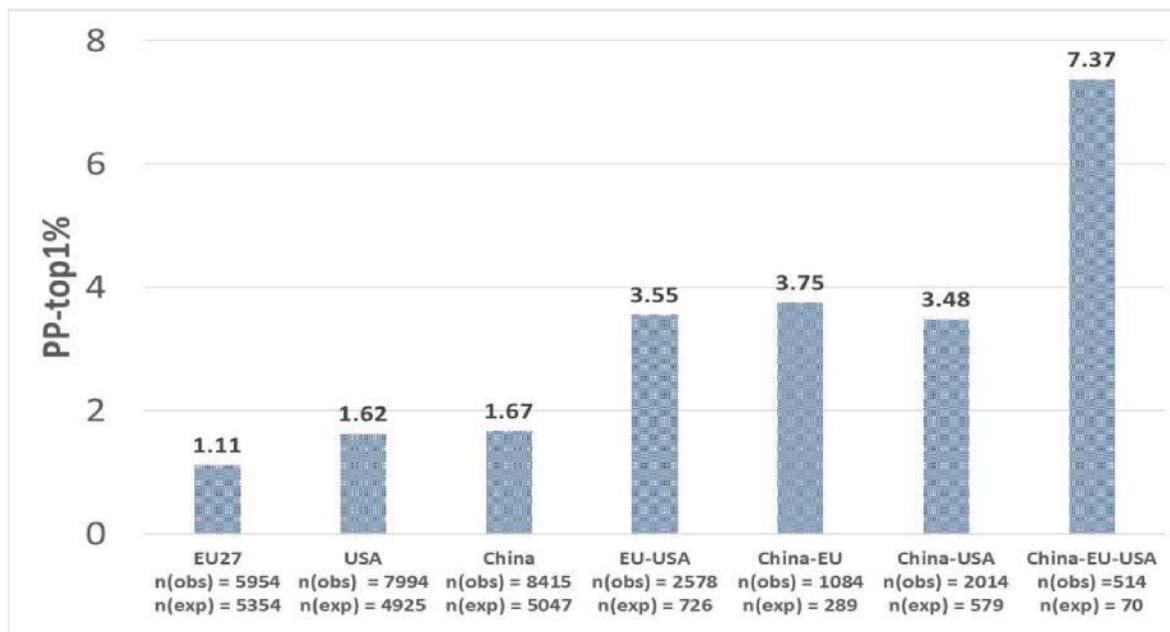

**Figure 4**: PP-top1% values for bi- and trilaterally co-authored papers (articles, reviews, and letters) published in 2019; downloaded from WoS on March 6, 2020, completed on April 24, 2021.

Figure 4 shows PP-top1% values for internationally co-authored (articles, reviews, and letters;) with at least one American, Chinese or European (EU27) address, respectively. The PP-top1% is almost doubled when two of the three blocks participate in a co-authorship, and doubles again for tripartite contributions. Here we see that each of the three entities alone score much lower than a combination of entities. International collaborative papers garner more attention and thus have higher above-expected citation rates.

*3.1. Whole-number and fractional counting*

Decisions regarding methods for counting addresses can also result in different outcomes when comparing nations. Gathered online, in our method, each nation is counted by a full count of 1 if an address appears in the bylines: called whole-number counting. "Fractional counting"—in which one divides credit according to the number of countries in the address byline—is used in science & technology policy analysis more than is whole-number counting. Fractional counting, however, lowers the count for international collaborations (Leydesdorff, 1988; Wagner, 2008; cf. Sivertsen *et al*., 2019). In our opinion, whole number counting can be considered as representing access to the knowledge content and participation in the respective knowledge bases, whereas fractional counting has advantages in research evaluations because all numbers then add up to 100% in a spreadsheet.

Using field-normalized and fractionally counted data published in Leydesdorff *et al*. (2014), Figure 5 shows the difference in this case for two overlapping years. As expected, whole-number counted data are associated with higher values of PP-top1%.

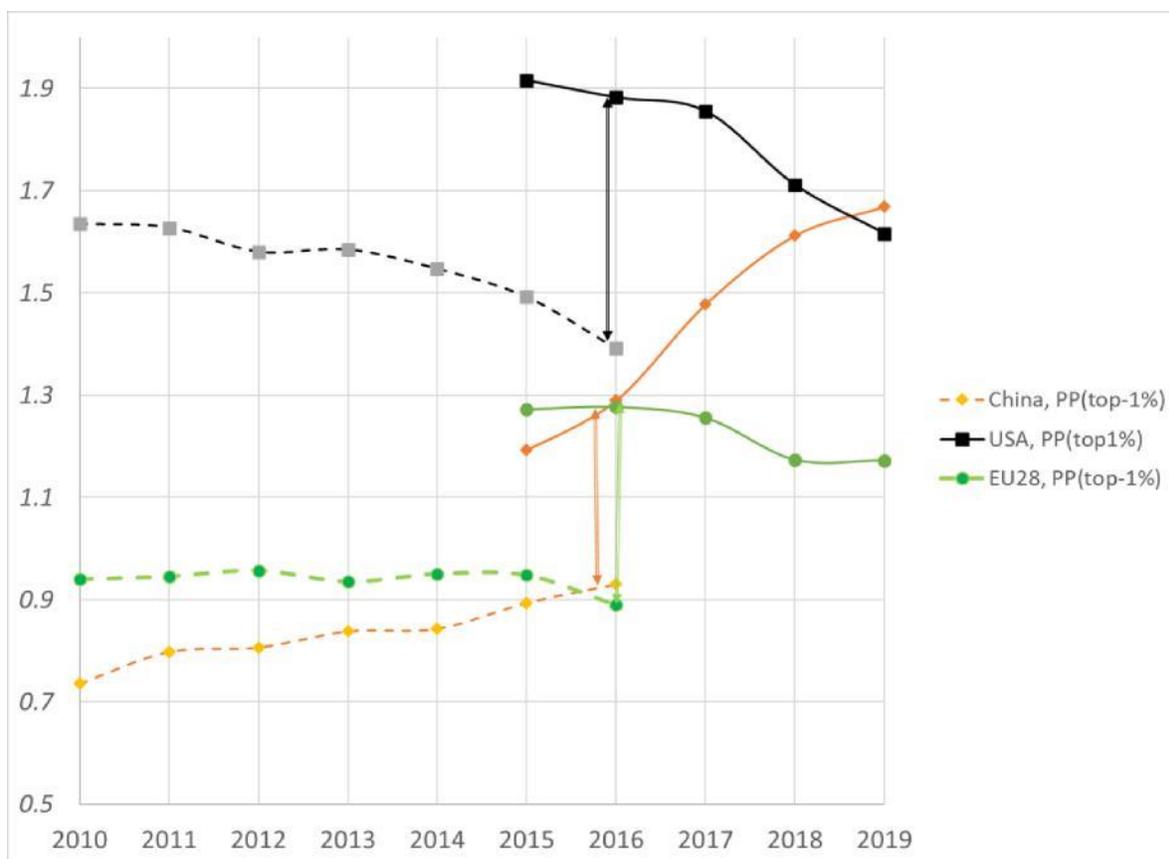

**Figure 5**. PP-top1% values for China, the United States, and the EU28 based on (1) fractionally counted (on the basis of countries) and field-normalized on the basis of WCs for the period 2010-2016 (dashed lines; source: Leydesdorff *et al*., 2014) and (2) whole-number counted for the period 2015-2019 (solid lines; source: this study).

Figure 5 shows that fractional counting, relative to whole-number counting, produces a lower value, the difference is greater for the USA than China.

*3.2. Citation windows*
We note that, using 2019 data, the citation window at the retrieval date (March 6, 2021) is short. The results for the earlier years (2015-2018) are more robust. However, we checked empirically that, in this case, the citation rates of these top-1% most-highly cited publications increase linearly with the citation window and therefore seem to be cited independently of age: while the top-1% were cited 38 or more times for 2019, this threshold was 67 for 2018, 93 for 2017, 115 for 2016, and 140 for 2015, respectively. The Pearson correlation between these values and the number of years in the respective citation windows is more than .99 (*p*<.01).

Because of this empirical finding, there was no need to apply a fixed citation window in this specific case.

**4. Decompositions into Subsets**

The analysis can be repeated using the PP-top1%—or other percentile ranks such as the top-10%—for subsets such as sets of journals attributed to a specific Web of Science subject category ("WC") or, for example, ESI "broad fields" or OECD field definitions (OECD, 2015). The observed values are measured as the participation in the pool of top-1% values after, for example, setting a filter such as "AND WC=virology." The expected top-1% values remain equal to the number of papers in the respective reference sets divided by 100.

**Table 2**: Comparison between China and the USA for four Web-of-Science Subject Categories in terms of shares in the 1% most-highly-cited publications in 2019. Data downloaded from Web-of-Science on March 6, 20121.

|  | N of journals | N of documents (ARL) | N of documents China | N of documents USA | P Top-1% China | P Top-1% USA | PP Top-1% China | PP Top-1% USA | $z$ (China:USA) |
|---|---|---|---|---|---|---|---|---|---|
| Virology | 37 | 6,625 | 1,387 | 2,480 | 13 | 41 | 0.94 | 1.65 | -1.805 |
| Engineering, biomedical | 87 | 13,365 | 2,952 | 3,923 | 54 | 61 | 1.83 | 1.55 | 0.878 |
| Engineering, multidisciplinary | 91 | 69,576 | 27,393 | 8,440 | 416 | 126 | 1.52 | 1.49 | 0.197 |
| Business and Finance | 109 | 6,048 | 961 | 2,157 | 15 | 16 | 1.56 | 0.74 | 2.133* |
| Without filters ("world") | (12,185) [5] | 2,041,287 | 504,695 | 492,448 | 7,959 | 8,442 | 1.67 | 1.62 | 0.804 |

* Significant at the 5% level.

---

[5] This is the number of journals in the *Journal Citation Reports* 2019.

To demonstrate decomposition at the field level, Table 2 shows the results derived from comparing four WoS Subject Categories from sciences, engineering, and social sciences. The fields "virology" and "business, finance" were chosen because these were relatively similar in number and represented relatively small sets ($n < 10,000$); the other two engineering fields (biomedical and multidisciplinary) were chosen because China tends to be strong in engineering, and thus these fields could offer useful comparisons.

Using PP-Top1%, the United States ranks ahead of China in biomedical fields such as virology. Although the number of papers is smaller, China outperforms the US in the field of Business and Finance (in the Social Sciences Citation Index); this difference is statistically significant at the 5% level. Analyses along these lines can compare nations to one another in terms of their disciplinary strengths and specializations.

Note that PP-top1% is a (size-independent) *quality* indicator. For the size-dependent *impact* one can multiply this value with the (relative) number of papers in the respective category and thus obtain the %*I3* (that is, the Integrated Impact Indicator; see Leydesdorff & Bornmann, 2011; Leydesdorff, Bornmann, & Adams, 2019; Wagner & Leydesdorff, 2012). *I3* is not a quality, but an impact indicator. Impact is size-dependent: *ceteris paribus* one expects two citations or publications, respectively, to generate more impact than a single one.[6]

Table 3 shows the quality indicator *PP-top1%* versus the impact indicator (*%I3*) for the two subsets—"virology" and "business, finance"—and the combination of the two subsets. The values in the numerical table 3 are visualized in Figure 6.

---

[6] One can formalize *I3* as follows:

$$I3 = \sum_{i=1}^{C} f(X_i) \cdot X_i \qquad (1)$$

where $X_i$ indicates the percentile ranks and $f(X_i)$ denotes the frequencies of the ranks with $i=[1,C]$ as the percentile rank classes.

**Table 3.** %PP-top1% and %/I3 for China and the US.

[Data downloaded as in Table 2, but on June 29, 2021.]

| Downloaded on June 29, 2021 | USA | China | with reference to the numbers in the respective full sets |
|---|---|---|---|
| **Virology** | 2463 | 1379 | 6622 |
| % Publications | 37.19 | 20.37 | |
| PP-tiop1/% | 1.38 | 0.8 | |
| *I3* | 444,624 | 183,984 | 1,486,371 |
| *%I3* | 29.91 | 12.38 | |
| **Business, finance** | 2160 | 961 | 6065 |
| % Publications | 35.61 | 15.85 | |
| PP-tiop1/% | 0.28 | 7.39 | |
| *I3* | 200,869 | 91909 | 742,532 |
| *%I3* | 27.05 | 12.37 | |
| **Combined** | 4,623 | 2,340 | 12,687 |
| % Publications | 36.44 | 18.44 | |
| PP-tiop1/% | 1.31 | 0.85 | |
| *I3* | 659,908 | 287,965 | 2,295,151 |
| %I3 | 28.75 | 12.54 | |
| | 492,461 | 504,373 | 2,040,618 |

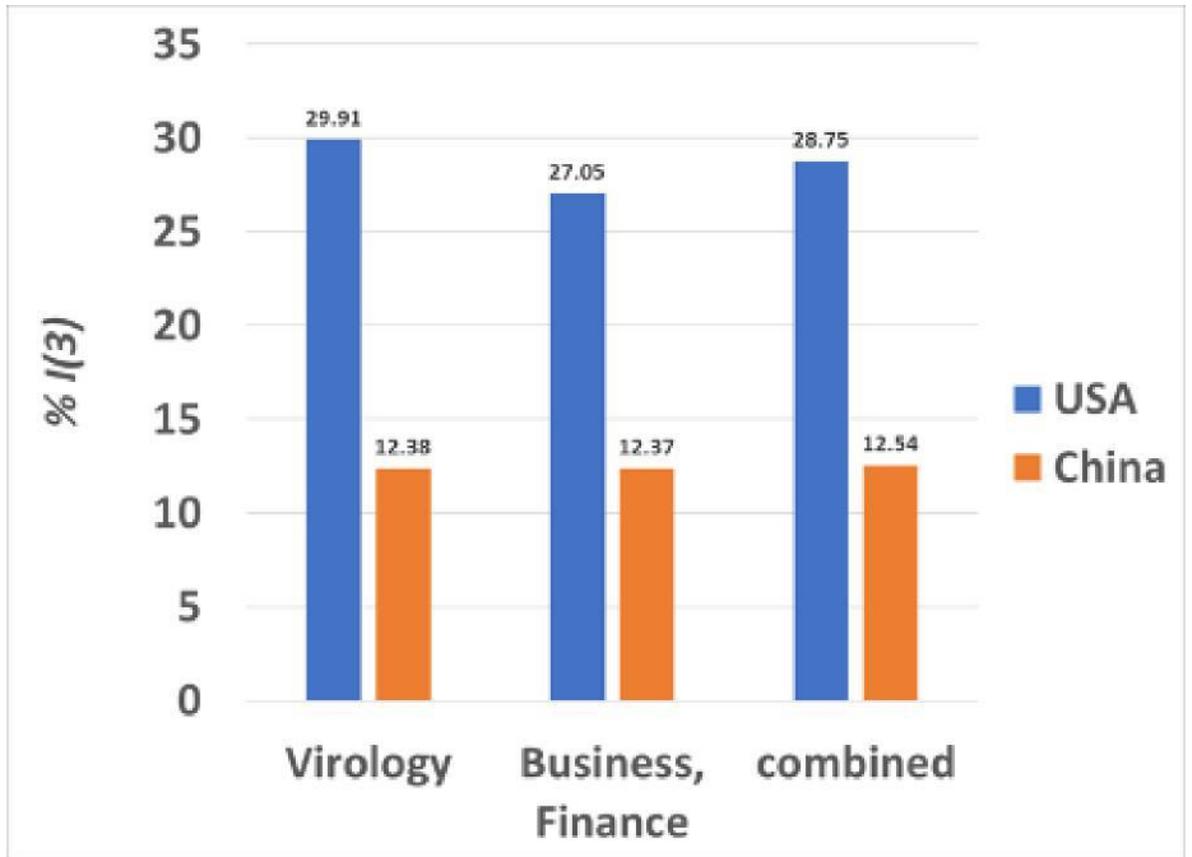

**Figure 6**: World shares of percentages of integrated impact (%*I*3) for articles, reviews, and letters within the domains of "virology" and "business, finance" and published in 2019.

In terms of impact (%*I3*), the US outperforms China, but the comparison can be very different in terms of quality (*PP-top1%*).

**5. Conclusions and policy implications**

Avoiding the use of field normalizations for calculating the top-1% most-highly cited may improve the validity of the measure. Using non-parametric statistics in calculating the set appears to be a measure more suited to the data. Using averages of citation rates within disciplines (in the denominator) distort the outcome. In our view, when calculating the top-1% most-highly cited at the national level, field-normalizations are not needed; percentile classes can be used instead. Drawing the top 1% of the set as the expectation, one can assess subsets beyond index-based classifications. Classifications can be local and can be used for policy analyses. Retrievals based on search strings or institutional units of analysis (e.g., universities in a nation) can analogously be compared with reference sets, for example (Bornmann & Leydesdorff, 2019). Either of the measures presented may disadvantage or advantage different fields based on the analyst'schoices, but it is important to avoid a statistical error introduced by using the mean in non-parametric statistics.

We note that the choice of indicators can have direct policy and funding impacts, thus, a less reliable indicator may point in the wrong direction. The top-1% analysis using field normalization may have obscured the fact that China is operating at world-leading levels of scientific output in both volume and quality. The increase of the *quality* of China's scientific output challenges a number of assumptions about the ways or conditions within which nations build scientific capacity. China's science policy has propelled the nation to world-class levels in a very short time period, moving the nation's profile from rapid imitation to levels challenging nations with a longer history of world-leading science.

## Acknowledgements

We thank Koen Jonkers, Xabier Goenaga, Ronald Rousseau and two anonymous referees for advice, critiques, and suggestions. Loet Leydesdorff is grateful to ISI/Clarivate for JR data.